\documentclass[a4paper,11pt]{article}
\usepackage{pos}

\usepackage[symbol]{footmisc}
\usepackage[english]{babel}
\usepackage{graphicx}
\usepackage{graphics}
\usepackage{braket}
\usepackage{bbold}
\usepackage{amsmath}
\usepackage{nicefrac}
\usepackage{dcolumn}
\usepackage{bm}
\usepackage{slashed}
\usepackage{datetime}
\usepackage{mciteplus}
\usepackage{multirow}
\usepackage{bbold}
\usepackage{siunitx}
\usepackage{booktabs}
\usepackage{color, soul}
\usepackage[usenames,dvipsnames]{xcolor}
\usepackage{float}
\usepackage[utf8]{inputenc}
\usepackage[normalem]{ulem}
\usepackage{mathtools}
\usepackage{setspace}

\renewcommand{\thefootnote}{\fnsymbol{footnote}}

\title{Proton 3D reconstruction with T-odd TMD gluon densities}
 \ShortTitle{Proton 3D reconstruction with T-odd TMD gluon densities}

\author[a,b]{Alessandro Bacchetta}
\author*[c]{Francesco Giovanni Celiberto}
\author[b]{Marco Radici}

\affiliation[a]{Dipartimento di Fisica, Universit\`a di Pavia, via Bassi 6, I-27100 Pavia}
\affiliation[b]{INFN Sezione di Pavia, via Bassi 6, I-27100 Pavia, Italy}
\affiliation[c]{Universidad de Alcalá (UAH), Departamento de Física y Matemáticas, Campus Universitario, Alcalá de Henares, E-28805, Madrid, Spain}

\emailAdd{alessandro.bacchetta@unipv.it}
\emailAdd{francesco.celiberto@uah.es}
\emailAdd{marco.radici@pv.infn.it}

\abstract{We present studies aimed at probing the 3D gluon content of the proton through spin-dependent TMD gluon densities, computed by means of the spectator-model approach. Our formalism incorporates a fit-based modulation function for the spectator mass, designed to capture longitudinal-momentum effects across a broad kinematic range. Special emphasis is placed on the time-reversal even Boer--Mulders and the time-reversal odd Sivers functions. Accurate understanding of these functions is crucial for conducting precise 3D analyses of nucleons, highlighting the importance of collaborative efforts between the LHC and EIC Communities.}

\FullConference{25th International Spin Physics Symposium (SPIN 2023)\\
 24-29 September 2023\\
 Durham, NC, USA\\}

\begin{document}
\maketitle

\section{Introduction}
\label{sec:introduction}

Exploring the internal structure of nucleons through a multi-dimensional analysis of their constituents represents a frontier in phenomenological studies at new-generation colliding machines.
The well-established collinear factorization, which relies on a one-dimensional description of the proton content based on collinear parton distribution functions (PDFs), has been successful in describing data at hadron-hadron and lepton-hadron colliders. However, fundamental questions regarding the dynamics of strong interactions remain unanswered. Understanding the origin of mass and spin of hadrons necessitates a paradigm extension from the purely collinear framework to a 3D tomographic treatment, which is naturally provided by the Transverse Momentum Dependent (TMD) formalism.

While our knowledge of quark TMD densities has seen significant progress in the last years, the gluon-TMD sector remains almost uncharted. 
A first list of (un)polarized gluon TMD PDFs at leading twist was given in Ref.~\cite{Mulders:2000sh} and then in Refs.~\cite{Meissner:2007rx,Lorce:2013pza,Boer:2016xqr}.
Phenomenological studies on unpolarized and polarized gluon TMD densities were proposed only recently~\cite{Lu:2016vqu,Lansberg:2017dzg,Gutierrez-Reyes:2019rug,Scarpa:2019fol,COMPASS:2017ezz,DAlesio:2017rzj,DAlesio:2018rnv,DAlesio:2019qpk}. 
A noteworthy distinction between collinear and TMD densities lies in the dependence on the gauge structure, with TMD PDFs being process-dependent due to their sensitivity to the transverse components of the gauge link~\cite{Brodsky:2002cx,Collins:2002kn,Ji:2002aa}.

In particular, quark TMD PDFs exhibit dependence on the process through the $[+]$ and $[-]$ \emph{staple} links. The gauge-link dependence of gluon TMDs is more intricate, involving two principal structures: the $f$-type (or Weisz\"acker--Williams) and the $d$-type (or dipole). The $f$-type and $d$-type densities are sensitive to $[\pm,\pm]$ future-pointing and $[\pm,\mp]$ past-pointing gauge-link combinations, respectively.
Beyond these two major structures, there exist more intricate gauge links. They are related to box-loop combinations of $[+]$ and $[-]$ staple links.
However, they are accessed via processes featuring multiple color exchanges between the initial and final state~\cite{Bomhof:2006dp}, thus leading to a (potential) violation of TMD factorization~\cite{Rogers:2013zha}.
A connection between the unpolarized gluon TMD function and the BFKL unintegrated distribution exists only at small-$x$ and moderate transverse momentum~\cite{Dominguez:2011wm,Hentschinski:2021lsh,Celiberto:2019slj} (see Refs.~\cite{Bolognino:2018rhb,Bolognino:2019pba,Motyka:2014lya,Brzeminski:2016lwh,Celiberto:2018muu,Bautista:2016xnp,Garcia:2019tne,Hentschinski:2020yfm,Boroun:2023goy,Luszczak:2022fkf,Cisek:2022yjj,Celiberto:2020wpk,Celiberto:2022grc,Celiberto:2016vhn,Bolognino:2018oth,Bolognino:2019yqj,Bolognino:2019yls,Celiberto:2020tmb,Celiberto:2020rxb,Bolognino:2021mrc,Celiberto:2021dzy,Celiberto:2021fdp,Celiberto:2022dyf,Celiberto:2015yba,Celiberto:2022gji,Celiberto:2022keu,Celiberto:2024omj,Celiberto:2022kxx,Celiberto:2023rzw} for recent applications).

Prime studies on quark TMD PDFs were conducted within the spectator-model formalism~\cite{Bacchetta:2008af,Bacchetta:2010si}, considering different spin states of di-quark spectators and various form factors for the nucleon-parton-spectator vertex. 
A pioneering extension to the T-even gluon TMD sector was done in Refs.~\cite{Bacchetta:2020vty} (see also Refs.~\cite{Bacchetta:2021oht,Celiberto:2021zww,Bacchetta:2021lvw,Bacchetta:2021twk,Celiberto:2022fam,Bacchetta:2023zir}).
In this work we give results for the T-even gluon Boer--Mulders function and a preliminary analysis of the $f$-type T-odd gluon Sivers TMD PDF, both calculated via an enchanced version of the spectator-model approach. 
The Boer--Mulders TMD can be accessed even in unpolarized collisions at the LHC, thus providing us with useful information about the distributions of linearly polarized gluons inside an unpolarized proton. 
The Sivers function is crucial for studying transverse single-spin asymmetries at future experiments, such as the Electron-Ion Collider (EIC) and the forthcoming LHCspin~\cite{Aidala:2019pit,Santimaria:2021uel,Passalacqua:2022jia}.

\section{Leading-twist gluon TMD PDFs in a spectator model}
\label{sec:TMDs}

We make use of an improved spectator-model approach to model the gluon TMD correlator. Within our assumption, a gluon featuring four-momentum $p$, transverse momentum $\boldsymbol{p}_T$, and longitudinal fraction $x$, is struck from a parent nucleon with four-momentum $P$ and mass $M$. 
Proton remainders are effectively treated as a single on-shell particle, referred to as the \emph{spectator}, possessing mass $M_X$ and spin-1/2.

The nucleon-gluon-spectator effective vertex incorporates two form factors, expressed as dipolar functions of $\boldsymbol{p}_T^2$
Our choice for nucleon-gluon-spectator vertex reads
\begin{equation}
 \label{eq:vertex}
 {\cal Y}^\mu_{ab} = \delta_{ab} \left( g_1(p^2)\gamma^\mu + g_2(p^2) \frac{i}{2M} \sigma^{\mu\nu}p_\nu \right) \, ,
\end{equation}
the $g_{1,2}$ functions being two dipolar functions of $\boldsymbol{p}_T^2$.
The choice of dipolar form factors dampens gluon-propagator singularities and mitigates logarithmic instabilities in the $|\boldsymbol{p}_T|$-integrated correlator.
All twist-two T-even gluon TMD PDFs in the proton were computed in~\cite{Bacchetta:2020vty}. In that study, the pure spectator approach was refined by weighting the spectator mass $M_X$ over a continuous range using a spectral modulation function designed to capture both small- and moderate-$x$ dynamics.
Its expression reads
\begin{equation}
\label{eq:rho}
 \rho^{\rm [spect.]} (M_X) = \mu^{2a} \left( \frac{A}{B + \mu^{2b}} + \frac{C}{\pi \sigma} e^{-\frac{(M_X - D)^2}{\sigma^2}} \right) \;.
\end{equation}
Thus, a given TMD PDF is written as
\begin{equation}
\label{eq:TMD}
 {\cal F}^g(x,\boldsymbol{p}_T^2) = \int_M^{\infty} d M_X \, \rho^{\rm [spect.]} (M_X) \, \hat{\cal F}^g(x, \boldsymbol{p}_T^2; M_X) \;,
\end{equation}
where $\hat{\cal F}^g$ stands for corresponding densities calculated in a pure spectator-model framework.

Values of parameters coming both from the couplings (Eq.~\eqref{eq:vertex}) and the modulation function (Eq.~\eqref{eq:TMD}) were fixed by performing a simultaneous fit of the $|\boldsymbol{p}_T|$-integrated unpolarized and helicity gluon TMD PDFs, $f_1^g$ and $g_1^g$, to the corresponding collinear densities provided by the {\tt NNPDF} Collaboration~\cite{Ball:2017otu,Nocera:2014gqa} at the initial energy scale of $Q_0 = 1.64$ GeV. The impact of statistical uncertainty was assessed using the bootstrap method~\cite{Forte:2002fg}.

\begin{figure}[t]
\centering

\includegraphics[width=0.4750\textwidth]{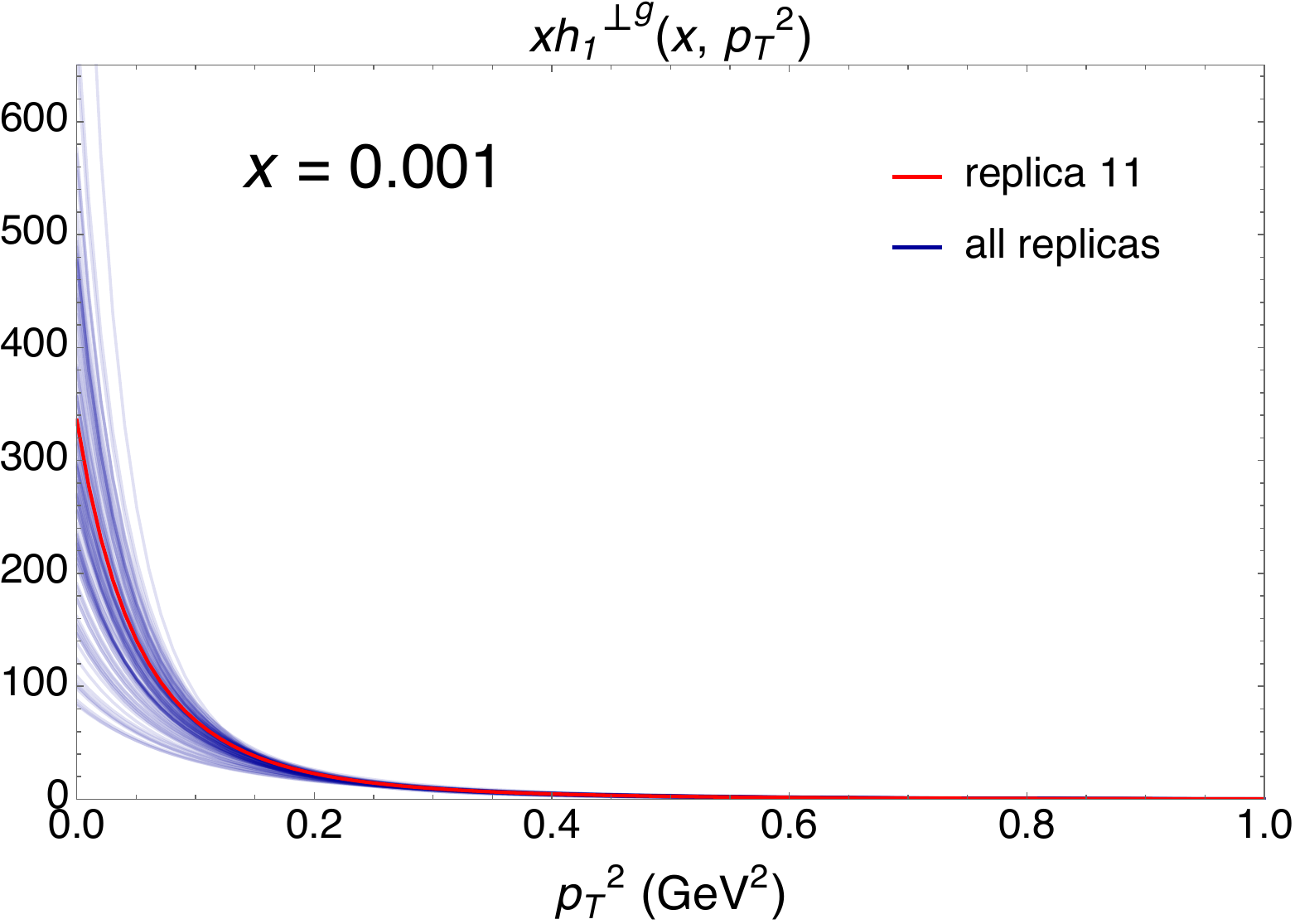} \hspace{0.40cm}
\includegraphics[width=0.4875\textwidth]{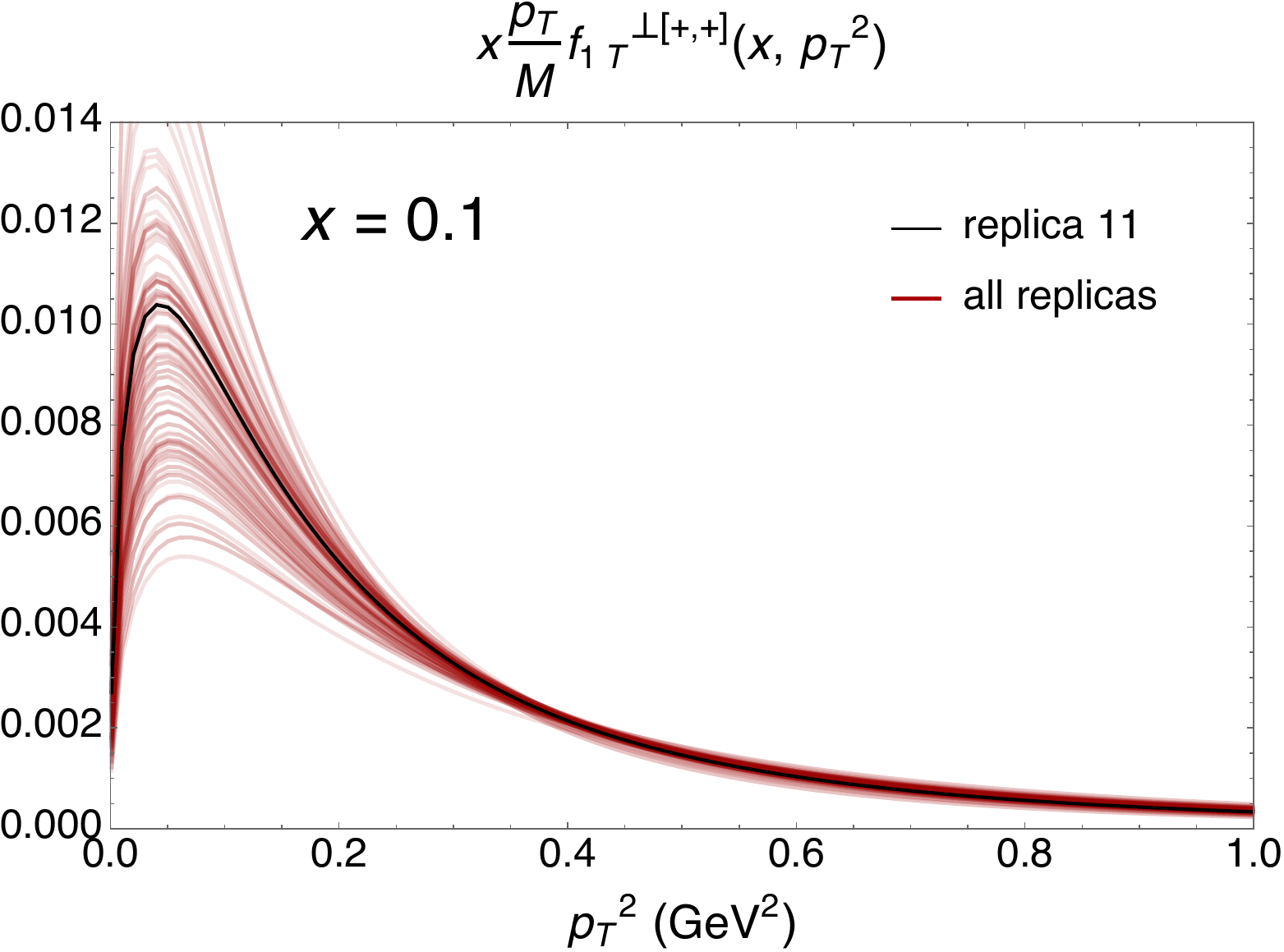}

\caption{Transverse-momentum trend of the Boer--Mulders gluon TMD at $x=10^{-3}$ (left) and the $|\boldsymbol{p}_T|$-weighted $[+,+]$ Sivers function at $x=10^{-1}$ (right), obtained within the spectator-model framework at the initial energy scale of $Q_0 = 1.64$ GeV. Black curve depicts replica \#11, which we elect as the most representative one.}

\label{fig:TMDs}
\end{figure}

The spectator-model gluon correlator at tree level carries no sensitivity on the gauge link. 
As consequence, our model T-even TMD PDFs do not depended on the process.
A T-odd function can be constructed by superseding the tree-level approximation and considering its interference with another channel. 
Using the same strategy employed for the quark TMD sector~\cite{Bacchetta:2008af}, we account for one-gluon exchange in the \emph{eikonal} approximation, representing the truncation at the first order of the full gauge-link operator. 
As a result, our T-odd densities acquire sensitivity on the gauge link.
Thus, they are process dependent.

A given spectator-model T-even TMD PDF, say the Boer--Mulders function ($h_1^{\perp g}$), does not change under a gauge-link modification.
At variance with the T-even case, two distinct T-odd distributions, such as two Sivers functions ($f_{1T}^{\perp g}$), appear. 
They can be obtained by suitably projecting the transverse part of the gluon TMD correlator. 
Thus, the following relations of \emph{modified universality} for the T-odd case hold
\begin{align}
 \label{eq:T-eodd_TMD_PDFs}
 f_{1T}^{\perp \, [+,+]}(x, \boldsymbol{p}_T^2) \; & = \; - \, f_{1T}^{\perp \, [-,-]}(x, \boldsymbol{p}_T^2) \; ,
 \; &[f\mbox{-tipe}]
 \\ \nonumber
 f_{1T}^{\perp \, [+,-]}(x, \boldsymbol{p}_T^2) \; & = \; - \, f_{1T}^{\perp \, [-,+]}(x, \boldsymbol{p}_T^2) \; .
 \; &[d\mbox{-tipe}]
\end{align}

We show results for the spectator-model gluon Boer--Mulders TMD PDF, along with preliminary analyses on the gluon Sivers function. 
For consistency, we use the model parameters those obtained from the fit of our integrated densities $f_1^g$ and $g_1^g$, given in Ref.~\cite{Bacchetta:2008af}, to {\tt NNPDF} collinear PDFs~\cite{Ball:2017otu,Nocera:2014gqa} at $Q_0 = 1.64$ GeV.
The panels in Fig.~\ref{fig:TMDs} illustrate the transverse-momentum dependence of the Boer--Mulders function (left) at $x = 10^{-3}$ and the $|\boldsymbol{p}_T|$-weighted $[+,+]$ Sivers distribution (right) at $x = 10^{-1}$. Both TMD PDFs feature a non-Gaussian $\boldsymbol{p}_T^2$ behavior, with a decreasing tail at high transverse momentum.
The Boer--Mulders function starts from a finite value at $\boldsymbol{p}_T^2 = 0$ and rapidly decreases as $\boldsymbol{p}_T^2$ increases. On the other hand, the Sivers function starts from a small, non-zero value at $\boldsymbol{p}_T^2 = 0$, then peaks in the range $\boldsymbol{p}_T^2 \lesssim 0.1$ GeV$^2$ before exhibiting a larger, flattening tail.

\section{Closing statements}
\label{sec:conclusions}

We presented a study of spin-dependent gluon twist-two TMD PDFs obtained within an improved version of the spectator-model approach~\cite{Bacchetta:2020vty}. 
The use of a fit-based modulation permitted us to effectively capture the core gluon dynamics at both small and moderate values of the longitudinal fraction $x$. 
Our ongoing efforts complete the calculation of all the T-odd gluon TMD PDFs~\cite{Bacchetta:2024fci}.

These investigations can provide valuable insights into the 3D dynamics of gluons inside the proton in the context of future colliding machines. 
These include: the Electron-Ion Collider (EIC)~\cite{AbdulKhalek:2021gbh,Khalek:2022bzd,Abir:2023fpo,Bolognino:2021niq}, NICA-SPD~\cite{Arbuzov:2020cqg}, the High-Luminosity Large Hadron Collider (HL-LHC)~\cite{Chapon:2020heu,Amoroso:2022eow,Begel:2022kwp,Bacchetta:2022nyv}, including its extension to polarized fixed targets~\cite{Lansberg:2012kf,Lansberg:2015lva,Kikola:2017hnp,Aidala:2019pit, Santimaria:2021uel,Passalacqua:2022jia}, and the Forward Physics Facility (FPF)~\cite{Anchordoqui:2021ghd,Feng:2022inv,Maciula:2022lzk,Celiberto:2022rfj,Celiberto:2022zdg,Bhattacharya:2023zei,Buonocore:2023kna}.

\section*{Acknowledgments}
\label{sec:acknowledgments}

This work is supported by the Atracci\'on de Talento Grant n. 2022-T1/TIC-24176, Comunidad Aut\'onoma de Madrid, Spain.

\begingroup
\bibliographystyle{bibstyle}
\bibliography{references}
\endgroup

\end{document}